# Reactivity of Stone-Wales defect in graphene lattice – DFT study

Aleksandar Z. Jovanović[1], Ana S. Dobrota[1], Natalia V. Skorodumova[2,3], Igor A. Pašti[1*]

[1] *University of Belgrade – Faculty of Physical Chemistry, Belgrade, Serbia*

[2] *Department of Materials Science and Engineering, School of Industrial Engineering and Management, KTH – Royal Institute of Technology, Stockholm, Sweden*

[3] *Applied Physics, Division of Materials Science, Department of Engineering Sciences and Mathematics, Luleå University of Technology, Sweden*

**Abstract**

Understanding the reactivity of carbon surfaces is crucial for the development of advanced functional materials. In this study, we systematically investigate the reactivity of graphene surfaces with the Stone-Wales (SW) defect using Density Functional Theory calculations. We explore the atomic adsorption of various elements, including rows 1-3 of the Periodic Table, potassium, calcium, and selected transition metals. Our results demonstrate that the SW defect enhances binding with the studied adsorbates when compared to pristine graphene, with carbon and silicon showing the most significant differences. Additionally, 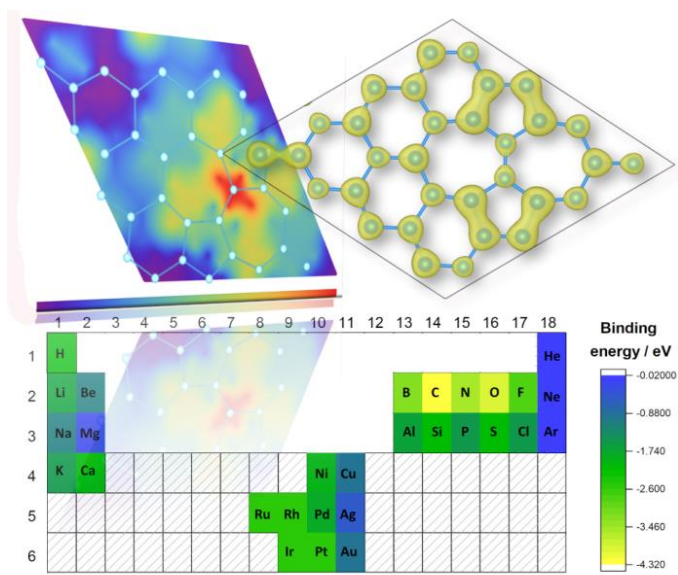 we examine the effects of mechanical deformation on the lattice by constraining the system with the SW defect to the pristine graphene cell. Interestingly, these constraints lead to even stronger binding interactions. Furthermore, for carbon, nitrogen, and oxygen adsorbates, we observe that mechanical deformation triggers the incorporation of adatoms into the carbon bond network, leading to the reorganization of the SW defect structure. This work establishes a foundation for future studies in the defect and strain engineering of graphene, opening avenues for developing advanced materials and catalysts with enhanced reactivity and performance.

---

[*] **corresponding author:** Prof. Igor A. Pašti, *University of Belgrade – Faculty of Physical Chemistry, Belgrade, Serbia*, E-mail: *igor@ffh.bg.ac.rs*

# 1. Introduction

Graphene is an allotrope of carbon consisting of a single layer of atoms arranged in a two-dimensional honeycomb lattice. It is the basic structural element of other C allotropes, including graphite, charcoal, carbon nanotubes, and fullerenes [1]. This atomic arrangement gives graphene its unique properties, including high electrical and thermal conductivity and extraordinary mechanical strength [2]. Due to these properties, graphene has been used in a variety of applications, including solar cells [3–5], touch screens [6], sensors [7,8], flexible electronics [9,10], energy storage [11–13], and many others. In fact, these days, it is very difficult to find any contemporary technology in which graphene is not present or at least tested for possible applications.

Although typically imagined as a flat sheet of atoms, in reality, graphene is not perfectly flat [14,15] and the presence of defects in the graphene structure cannot be avoided. Various types of defects can occur in graphene, broadly classified into point defects and line defects [16,17]. Point defects are localized disruptions to the regular atomic arrangement and can be either vacancies (missing atoms) or impurities (substitutions of foreign atoms) [17,18]. Line defects are linear disruptions to the atomic structure and include things like edge dislocations and grain boundaries [17,18]. They can be created by removing or substituting multiple carbon atoms or introducing a foreign atom into the lattice.

Both point and line defects can have a significant impact on the properties of graphene [17]. For example, point defects can act as sites for chemical reactions [17] even if no dangling bonds are present, while line defects can affect the electrical and thermal conductivity of graphene [18,19]. Therefore, it is important to carefully control the number and type of defects present in graphene when it is being used for applications such as electronics or energy storage [20,21]. In general, the more defects there are in a graphene sample, the more its properties will be degraded with respect to that of pristine graphene. However, the presence of defects is crucial for some applications of graphene-based materials. This particularly relates to chemical applications, including chemical reactions with or on the graphene surface, as pristine graphene is exceptionally chemically inert and interacts very weakly with most atomic adsorbates [22]. However, vacancies are the sites of enhanced reactivity and interact with different species much stronger than pristine graphene [23], resulting from the formation of localized states near the vacancy.

Another common type of defect in graphene is the Stone-Wales (SW) defect, which is also present in other nanocarbon materials [17]. The SW defect results from rotating two carbon atoms about a bond axis, resulting in two five-membered rings and two seven-membered rings instead of four hexagons [24]. For this reason, the SW defect is also called the 5775 defect, and its formation takes approximately 5 eV [25,26]. Like in the case of vacancies in graphene, the presence of Stone-Wales defects can significantly impact the electronic structure, chemical reactivity, and adsorption properties of the material [27–29], acting as adsorption sites for molecules or ions or increasing the rate of chemical reactions by providing an active site for the reaction [30,31]. For example, the presence of SW defect enhances the interaction of graphene with ambient gases ($N_2$, $O_2$, Ar, $CO_2$, and $H_2O$), causing damage in the material, particularly in the presence of oxygen [30]. However, it can also be utilized to turn graphene sheets into a membrane with a high proton permeability and isotope selectivity [31].

Besides the alteration of the chemical properties of graphene due to the presence of defects, mechanical deformations can significantly impact the reactivity of graphene [32]. Thus, it is essential to consider how mechanical deformations will affect the reactivity of graphene before using it in applications where it may be subject to such forces. For example, reactivity tuning by mechanical strain could be employed to effectively store and release hydrogen from the graphene surface [33–35]. In contrast, combining heteroatoms and mechanical strain can induce side-selective reactivity towards different species [36].

In our previous works, we have addressed atomic adsorption on pristine graphene [22] and graphene with single vacancy [23] in a systematic fashion, considering all the elements in the first six rows of the Periodic Table of Elements (PTE). In this work, we continue the quest for understanding the reactivity trends of graphene surfaces by analyzing the atomic adsorption of graphene with SW defect. We have restricted our study to the elements located in rows 1-3 of the PTE, additionally including K, Ca, d-elements, which are particularly interesting for catalysis and electrocatalysis (Ni, Ru, Rh, Pd, Ir, Pt), and coinage metals (Cu, Ag, and Au). In the present work, we have also analyzed how the selected elements interact with the SW defect in the lattice under compressive strain invoking structural deformation of the surface layer. The presented results provide insight into the reactivity trends of the SW-containing graphene and can help develop novel strategies for graphene functionalization by defect and strain engineering.

## 2. Computational details

We calculated the adsorption of all the elements of the PTE located in rows 1 to 3, in the 4×4 cell of graphene (32 atoms) containing a SW defect. Additionally, we analyzed the adsorption of K, Ca, selected d-elements (Ni, Ru, Rh, Pd, Ir, Pt), and coinage metals (Cu, Ag, and Au). Due to the systematic nature of this work, the minimal simulation cell size (32 atoms), which does not induce strong interactions between periodic images, was used.

The first-principle DFT calculations were performed using the Vienna *ab initio* simulation code (VASP) [37–40]. We used the generalized gradient approximation (GGA) in the parametrization by Perdew, Burk and Ernzerhof [41] and the projector augmented wave (PAW) method [42,43]. To account for dispersion interactions, the DFT+D3 formulation of Grimme was used [44]. In this approach, the total energy is corrected by a pairwise term, which accounts for dispersion interactions and is added to the total energy of the system calculated using PBE functional. The cut-off energy of 600 eV and Gaussian smearing with the width of $\sigma$ = 0.025 eV for the occupation of the electronic levels were used. A Monkhorst-Pack $\Gamma$-centered 10×10×1 k-point mesh was used. We ~~have~~ selected 31 initial adsorption sites (**Figure S1**, Supplementary Information), which were systematically investigated. Two sets of calculations were done. First, the calculations were done with the cell corresponding to that of pristine graphene. As explained later on, this induces the corrugation of the SW-containing lattice due to compressive strain. In this set of calculations, all the atoms in the cell were allowed to fully relax. In the second set of calculations, the atoms and the cell were fully relaxed. The relaxation procedure was stopped when the Hellmann-Feynman forces on all atoms were smaller than $10^{-2}$ eV Å$^{-1}$. This corresponded to the total energy converged below 0.01 meV. Spin-polarization was taken into account in all calculations. The repeated graphene sheets were separated from each other by 20 Å of vacuum.

The binding energies ($E_b$) obtained within different sets of calculations were calculated as:

$$E_b^{rel}(A) = E_0^{rel}(A@SW-G) - E_0^{rel}(SW-G) - E_0(A) \qquad (1)$$

$$E_b^{pris}(A) = E_0^{pris}(A@SW-G) - E_0^{pris}(SW-G) - E_0(A) \qquad (2)$$

where $E_0$ are the ground state energies of the adatom on SW-graphene [A@SW–G], SW-graphene [SW–G], and adatom [A] alone. Superscript "rel" or "pris" indicates whether the supercell was allowed to relax or was fixed to that of a pristine graphene lattice. $E_b$ is negative when adsorption is exothermic.

Visualization was done using VESTA [45], while the graphical presentation of Densities of States (DOS) was done using sumo tools for VASP [46].

## 3. Results and discussion

3.1. SW defect in the graphene lattice

As mentioned in the introduction, the SW defect is formed by a rotation of one C–C bond, forming two five-membered and two seven-membered rings. This process requires very high energy input. Hence, the formation of the SW defect is not probable under low temperatures [17,25]. However, if formed during the synthesis and quenched in the structure, a high energy barrier has to be overcome to restore the perfect lattice. Thus, it remains in the structure.

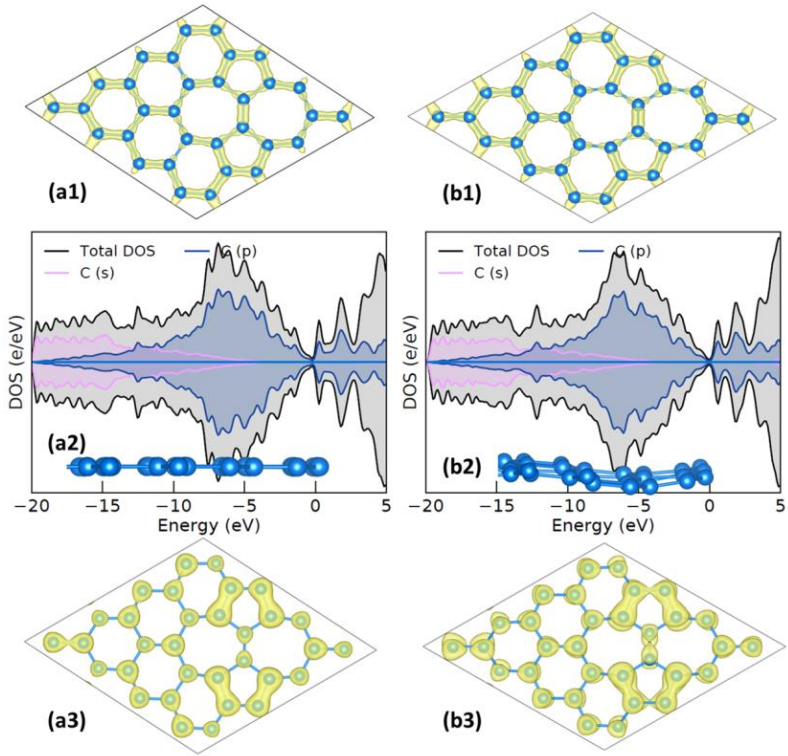

**Figure 1.** Two structures of the SW defect in graphene lattice. On the left (a), the SW defect in a fully relaxed lattice is presented with charge density distribution, Densities of States (DOS), and isosurfaces of partial charge density obtained by the DOS integration from –2 eV do Fermi level (set to 0). On the right (b), analogous data for the SW defect constrained to the pristine graphene lattice are presented.

Here we introduced the SW defect into the (4×4) supercell of pristine graphene and relaxed the structure in two ways. First, the cell and all the atoms were fully relaxed while keeping the simulation cell volume constant. This resulted in an almost perfectly flat graphene sheet with the embedded SW defect (**Figure 1**, a). This finding agrees with previous findings, where only minor deviations from the planarity were reported for the SW defect in the sp$^2$-bonded planar materials [26]. In the second approach, we allowed only ionic relaxation while maintaining the cell to that of pristine graphene. As a result, the graphene sheet has undergone deformation, resulting in the development of curvature (**Figure 1**, b). The electronic structures of the obtained SW-graphene surface models are quite similar (**Figure 1**, a1 and a2), although we have observed a small build-up of the electronic states in the vicinity of the Fermi level for the case of the system where the cell was fixed. **Figure 1**, a3 and b3, compares partial charge densities obtained by the density of states (DOS) integration from –2 eV to Fermi level (0 eV) for the two systems considered. This DOS range was selected as it was previously shown that it could be linked with the reactivity of graphene [47]. While partial charge densities are qualitatively similar, less charge was found in the case of a fully relaxed system. For the fully relaxed system, we found 4.8 electrons in the −2 to 0 eV window, while for the system in the pristine graphene cell, we found 5.18 electrons in total (per 32 atoms). The largest difference is observed for the carbon atoms forming the SW defect, thus, it is expected that the charge accumulation caused by the mechanical deformation of the sheet will additionally increase the reactivity of the SW defect in the graphene lattice.

To study atomic adsorption on the SW-containing graphene plane, we adopted a similar approach in terms of structural relaxation. Thus, the systems where the cell and ionic relaxation were allowed are denoted as "Relaxed cell", while the systems where only ionic relaxation was allowed were denoted as "Pristine graphene cell". Next, we discuss atomic adsorption with respect to the adsorbate location in the PTE. First, hydrogen and s- metals are discussed. Then, we move to the p-elements and continue with noble gases. Finally, we discuss transition metals with relevance to catalysis. Finally, the reactivity of the SW-defective graphene is compared to that of pristine graphene to provide a complete overview of the reactivity trends.

3.2. H and s-metals

For the group 1 elements, the binding energies decrease from H to Na, with the exception of K, having the highest $E_b$ in the series (**Table 1**). This trend suggests that the binding strength of Group 1 elements to graphene with a Stone-Wales defect decreases as moving down the group, likely due to the increasing atomic size and decreasing electronegativity along the group. Hydrogen forms a single H–C bond (on top adsorption; **Figure 2**), while Li, Na, and K position themselves in the center of the seven-membered ring. These geometries were found for both relaxed cell calculations (**Figure 2**) and the calculations in the pristine graphene cell (**Figure S2**, Supplementary Information).

The binding energy of Be is lower than that of Mg, while Ca has a significantly lower binding energy compared to the other two elements. When the cell is fully relaxed, Be is accommodated in the center of the five-membered ring of the SW defect (**Figure 2**). However, when the cell is constrained, it binds directly to the C atom, forming on Be–C bond (**Figure S2**, Supplementary Information). Mg is located very far from the surface, while Ca adsorbs in the center of the seven-membered ring (**Figure 2** and **Figure S2**, Supplementary Information).

When comparing atomic adsorption on a fully relaxed surface and the constrained one, modifications of reactivity can be observed. H, Be, and Ca show enhanced binding on constrained surfaces (pristine graphene cell), while Li and Na show weaker bonding on the constrained surface. Finally, Na and Mg adsorption is weakly affected by surface mechanical deformation (**Table 1**).

**Table 1.** Binding energies ($E_b$), adsorbate-C nearest distance ($d$), and total magnetization (M) for adsorption of hydrogen, alkaline, and alkaline earth elements.

|  | Relaxed cell | | | Pristine graphene cell | | |
| --- | --- | --- | --- | --- | --- | --- |
| **Adsorbate** | $E_b$ / eV | $d$ / Å | $M$ / $\mu_B$ | $E_b$ / eV | $d$ / Å | $M$ / $\mu_B$ |
| **H** | −1.836 | 1.113 | 0.00 | −2.121 | 1.108 | 0.00 |
| **Li** | −1.732 | 2.187 | 0.00 | −1.649 | 2.180 | 0.00 |
| **Na** | −1.136 | 2.571 | 0.00 | −1.079 | 2.568 | 0.00 |
| **K** | −1.575 | 2.890 | 0.00 | −1.581 | 2.795 | 0.00 |
| **Be** | −0.902 | 1.904 | 0.98 | −1.100 | 1.730 | 0.00 |
| **Mg** | −0.241 | 3.727 | 0.00 | −0.272 | 3.368 | 0.00 |
| **Ca** | −1.821 | 2.417 | 0.00 | −2.113 | 2.346 | 0.00 |

Considering the overall trends, the electronic structure (see DOS plots, **Figure 2** and **Figure S2**, Supplementary Information) is not significantly affected by the adsorption of H and s-metals on the SW-defective graphene. Some modifications can be observed while Fermi level shifts are the consequence of pronounced charge transfer, that is, n-doping when s-metals are adsorbed on the surface. Moreover, magnetization was observed only for Be in a fully relaxed cell (**Table 1**).

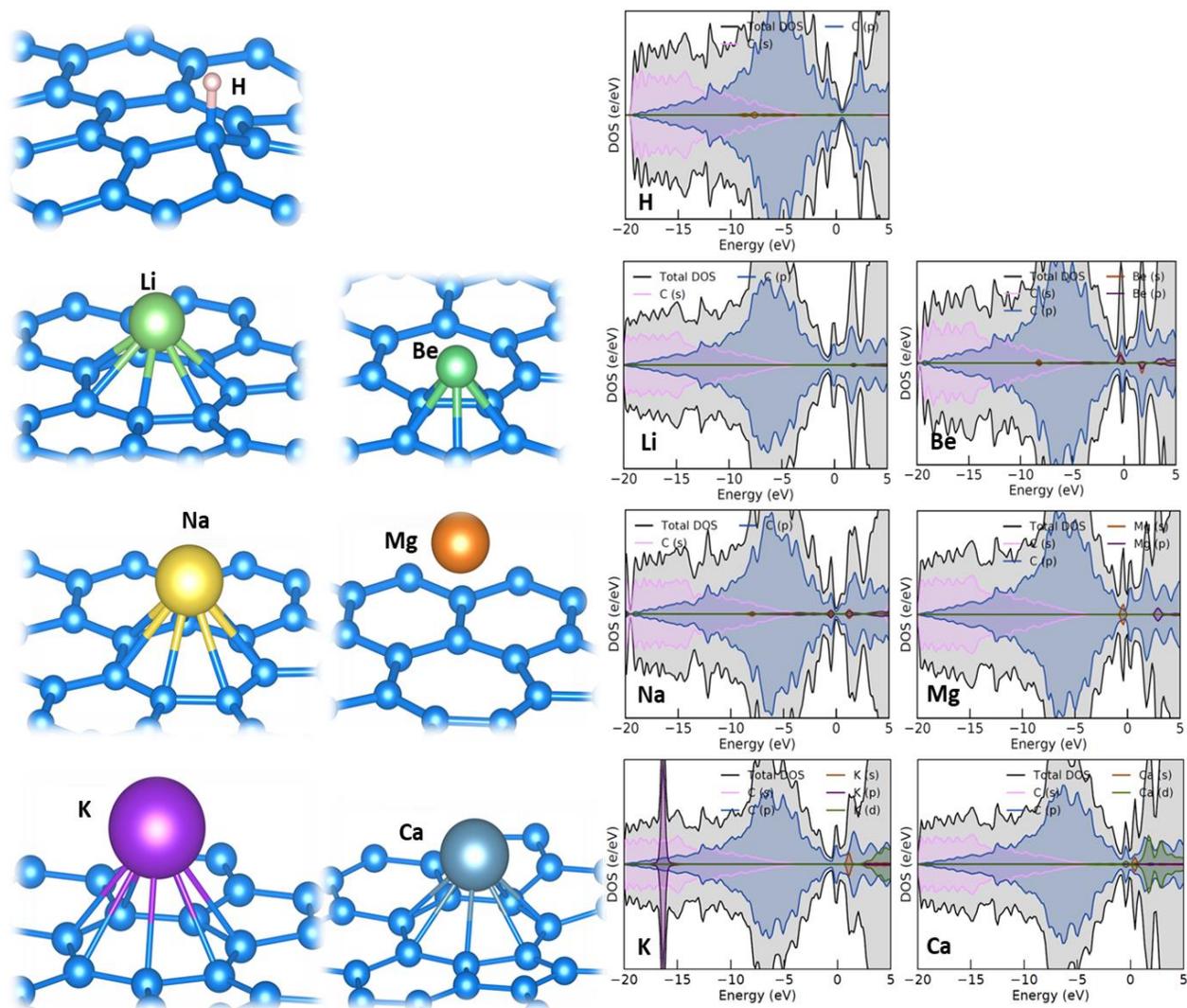

**Figure 2.** Optimized structures and DOSes for H, alkaline, and alkaline earth elements interacting with SW-graphene in fully relaxed lattice.

3.3. p-elements

The binding energies of the p-elements on the relaxed cell of graphene with the Stone-Wales defect are generally more negative compared to the binding energies of Group 1 and 2 elements. This results from forming covalent bonds between the adsorbates and the surface (**Figure 3**). Namely, when the cell is fully relaxed, B adsorbs at the center of the five-membered ring of the SW defect, while Group 4, 5, and 6 elements bind at the bridge position. Halogens (F and Cl) bind directly on top of the C atom in the center of the defect. Across the p-elements, there is a general decreasing trend in binding energies from left to right within the periodic table (B to Cl). However, $E_b$ passes through the maximum for the elements with $ns^2np^3$ configuration (N and P) (**Table 2**).

When comparing the binding energies of the p-elements in the relaxed cell calculations to those in the constrained cell calculations, we can observe more pronounced changes in the binding energies compared to the Group 1 and 2 elements. In general, except for Al, the binding energies of the p-elements on the constrained lattice of pristine graphene become more negative (more favorable) compared to the relaxed cell calculations. This finding suggests that mechanical deformation enhances the binding strength of the p-elements to the graphene surface with the Stone-Wales defect. The overall trends in binding energies are similar for the relaxed cell and constrained cell calculations.

It is interesting to observe the distinct behavior of C, N, and O adsorbates in the fully relaxed (**Figure 3**) and the constrained cells (**Figure S3**, Supplementary Information). In the fully relaxed cell, these atomic adsorbates bind at the bridge site between the two central carbon atoms of the SW defect. However, if the cell is constrained (and exposed to mechanical deformation), these adsorbates get incorporated into the lattice, breaking the SW defect and forming a defective structure which could be denoted as "578" in analogy to the "5775" notation of the SW defect. Such a behavior aligns with the previous computational predictions that C adatoms can heal the SW defect in graphene lattice [48] by opening the defect structure and recombining it to the pristine one with the activation energy of only 0.87 eV (which is still too high to take place at the room temperature). Thus, here we find that similar behavior might be expected for N and O. At the same time, the SW defect reconstruction could also be triggered by the mechanical deformation of the graphene sheet.

**Table 2.** Binding energies ($E_b$), adsorbate-C nearest distance ($d$), total magnetization ($M$) for p-elements in rows 2 and 3 of the PTE.

| | Relaxed cell | | | Pristine graphene cell | | |
|---|---|---|---|---|---|---|
| Adsorbate | $E_b$ / eV | $d$ / Å | $M$ / $\mu_B$ | $E_b$ / eV | $d$ / Å | $M$ / $\mu_B$ |
| B | −1.917 | 1.932 | 0.00 | −3.193 | 1.806 | 0.00 |
| C | −2.877 | 1.533 | 0.00 | −4.301 | 1.395 | 0.00 |
| N | −2.121 | 1.464 | 0.85 | −3.569 | 1.336 | 0.00 |
| O | −3.053 | 1.447 | 0.00 | −3.996 | 1.386 | 0.00 |
| F | −2.592 | 1.460 | 0.00 | −2.878 | 1.440 | 0.00 |
| | | | | | | |
| Al | −1.654 | 2.388 | 0.01 | −1.604 | 2.483 | 0.00 |
| Si | −1.958 | 1.976 | 0.00 | −2.080 | 1.942 | 0.00 |
| P | −1.343 | 1.907 | 0.94 | −1.537 | 1.886 | 0.95 |
| S | −1.848 | 1.861 | 0.00 | −2.068 | 1.840 | 0.00 |
| Cl | −1.269 | 1.957 | 0.00 | −1.521 | 1.889 | 0.00 |

Due to the formation of chemical bonds between the defective graphene surface and the adsorbates, the electronic structure is more disrupted (**Figure 3** and **Figure S3**, Supplementary Information) compared to the case of Group 1 and 2 elements adsorption, but we have not seen any bandgap opening. However, in the case of P adsorption, we observe magnetization in both cells considered here (**Figure 4**). The magnetic moment is primarily located on the P adatom. In the case of N, the magnetization is observed only in the case of the fully relaxed cell (**Table 2**) when N is adsorbed at the bridge site between the central carbon atoms of the SW defect. When the cell is constrained, N ignites the reconstruction of the SW defect, and magnetization disappears.

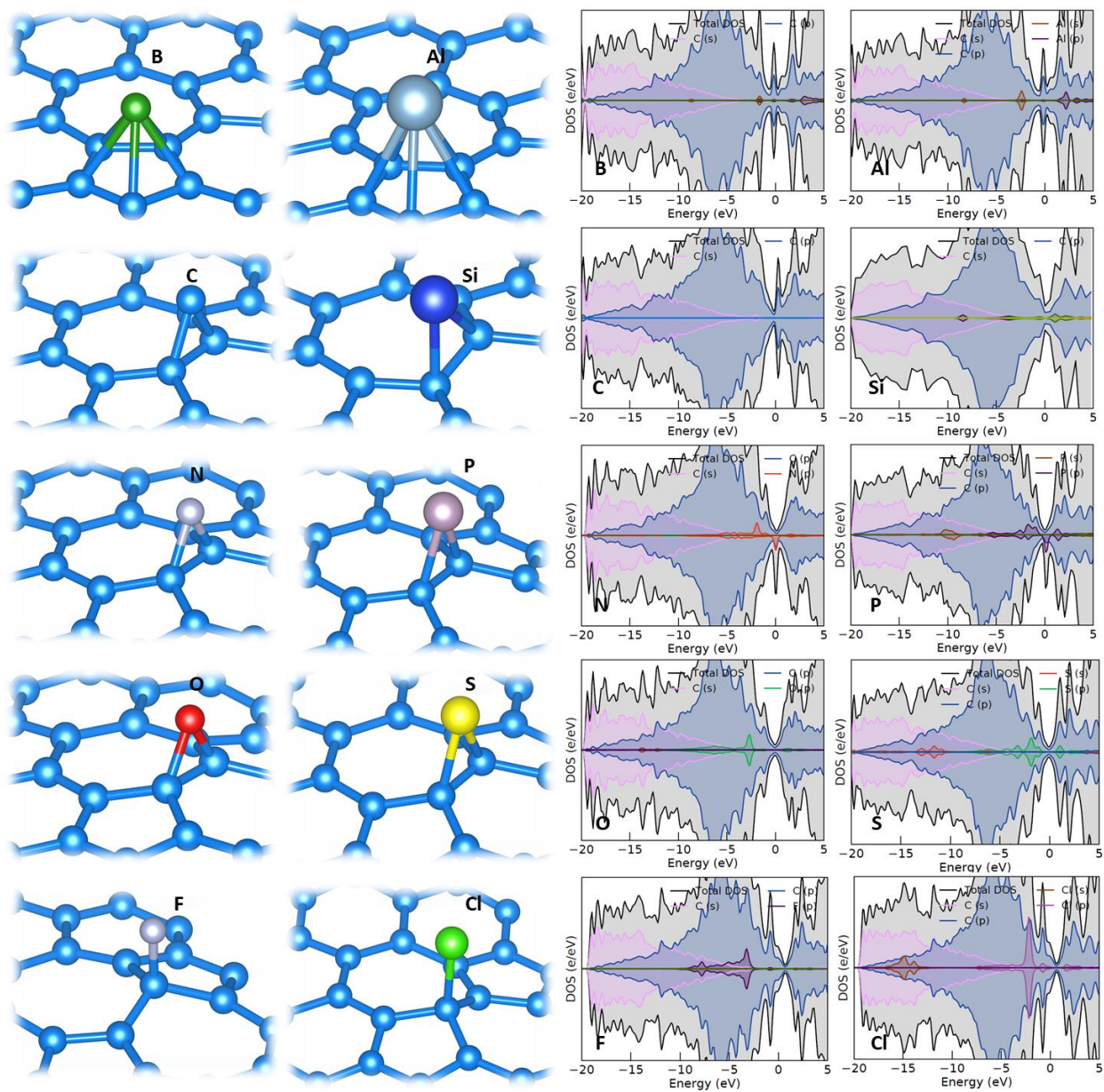

**Figure 3.** Optimized structures and DOSes for p-elements in rows 2 and 3 of the PTE at the SW defect in a fully relaxed cell.

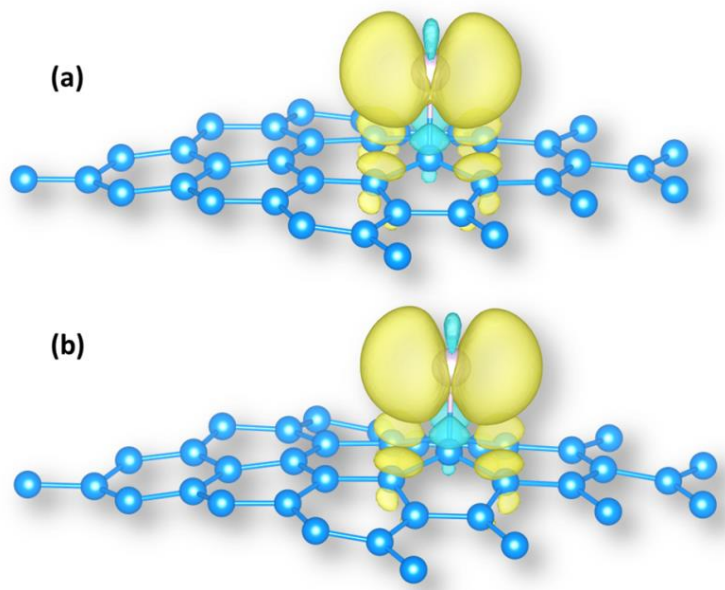

**Figure 4.** The spin density of the N@SW-graphene system (isosurface values 1.5×10⁻³ e A⁻³): (a) fully relaxed cell, (b) pristine graphene cell.

3.4. Noble gasses

Noble gases, as expected, interact very weakly with the SW-defective graphene surface. There are slight differences in the binding energies when comparing the relaxed cell and constrained cell calculations. In general, the fixed cell calculations result in slightly stronger binding compared to the relaxed cell calculations for all the considered noble gases (**Table 3**). Irrespective of the cell relaxation, the noble atoms are located more than 3 Å away from the surface (**Table 3**, **Figure 5**, and **Figure S4**, Supplementary Information).

Among the noble gases considered, the binding energies increase from helium to neon to argon, indicating a weak trend of increasing binding strength. This trend is the same as for the case of noble gas adsorption on pristine graphene, where the binding energies were found to scale with the polarizabilities of the noble gas atoms [22].

**Table 3.** Binding energies ($E_b$), adsorbate-C nearest distance ($d$), total magnetization ($M$) for noble gases interacting with SW-graphene.

|  | Relaxed cell | | | Pristine graphene cell | | |
| --- | --- | --- | --- | --- | --- | --- |
| **Adsorbate** | $E_b$ / eV | $d$ / Å | $M$ / $\mu_B$ | $E_b$ / eV | $d$ / Å | $M$ / $\mu_B$ |
| **He** | −0.029 | 3.166 | 0.00 | −0.034 | 3.076 | 0.00 |
| **Ne** | −0.046 | 3.334 | 0.00 | −0.053 | 3.221 | 0.00 |
| **Ar** | −0.096 | 3.693 | 0.00 | −0.106 | 3.446 | 0.00 |

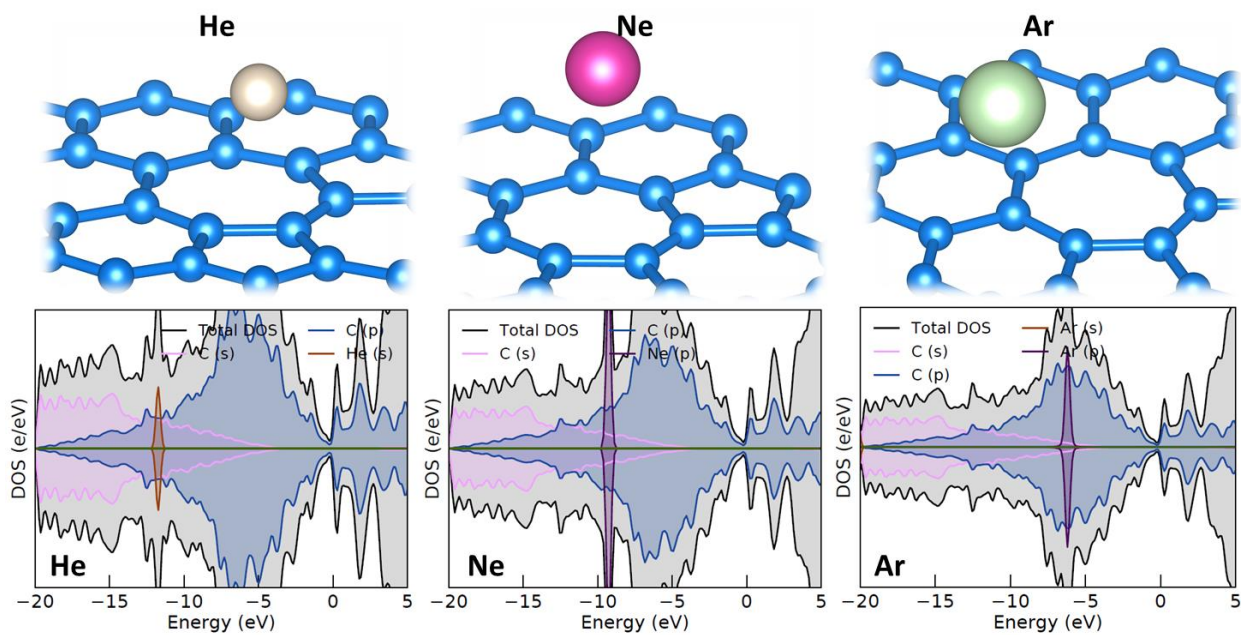

**Figure 5.** Optimized structures and DOSes of noble gases interacting with the SW defect in a fully relaxed supercell.

### 3.5. d- and coinage metals

For all of the d- and coinage metals, the binding energies in the pristine graphene lattice configuration are slightly more negative compared to those for the relaxed cell configuration (**Table 4**). This result suggests that the mechanical deformation of the graphene surface with the Stone-Wales defect can enhance the binding affinity of the transition and coinage metals. Particular trends along the row and the group depend on the actual electron configuration of a given element. However, there is a clear trend of more negative binding energies for the considered transition metals (Ni, Ru, Rh, Pd, Ir, Pt) compared to the coinage metals (Cu, Ag, Au). We find that the constraining surface relaxation affects the final geometry of adsorbed metals. In the fully relaxed cell, Ru and Rh form a multicentric bond, being adsorbed at the center of the five-membered ring (**Figure 6**). In contrast, when the cell is constrained (**Figure S5**, Supplementary Information), Ru binds directly on top of the C atom in the center of the SW defect, while Rh prefers bridge configuration. Irrespective of the cell relaxation, Ni, Pd, Ir, and Pt adsorb in the bridge configuration in the center of the SW defect, while coinage metals prefer on-top adsorption (**Figure 6**).

**Table 4.** Binding energies ($E_b$), adsorbate-C nearest distance ($d$), total magnetization ($M$) of selected d-elements and coinage metals.

| | Relaxed cell | | | Pristine graphene cell | | |
|---|---|---|---|---|---|---|
| **Adsorbate** | $E_b$ / eV | $d$ / Å | $M$ / $\mu_B$ | $E_b$ / eV | $d$ / Å | $M$ / $\mu_B$ |
| **Ni** | −1.881 | 1.918 | 0.00 | −1.906 | 1.903 | 0.00 |
| **Cu** | −0.767 | 1.993 | 0.47 | −0.927 | 1.949 | 0.02 |
| **Ru** | −2.315 | 2.069 | 0.00 | −2.467 | 2.014 | 1.08 |
| **Rh** | −2.485 | 2.127 | 0.00 | −2.522 | 2.028 | 0.28 |
| **Pd** | −1.738 | 2.123 | 0.00 | −1.797 | 2.092 | 0.00 |
| **Ag** | −0.303 | 2.338 | 0.36 | −0.451 | 2.255 | 0.00 |
| | | | | | | |
| **Ir** | −2.366 | 2.048 | 0.82 | −2.463 | 2.032 | 0.84 |
| **Pt** | −2.433 | 2.056 | 0.00 | −2.542 | 2.036 | 0.00 |
| **Au** | −0.767 | 2.149 | 0.00 | −0.981 | 2.103 | 0.00 |

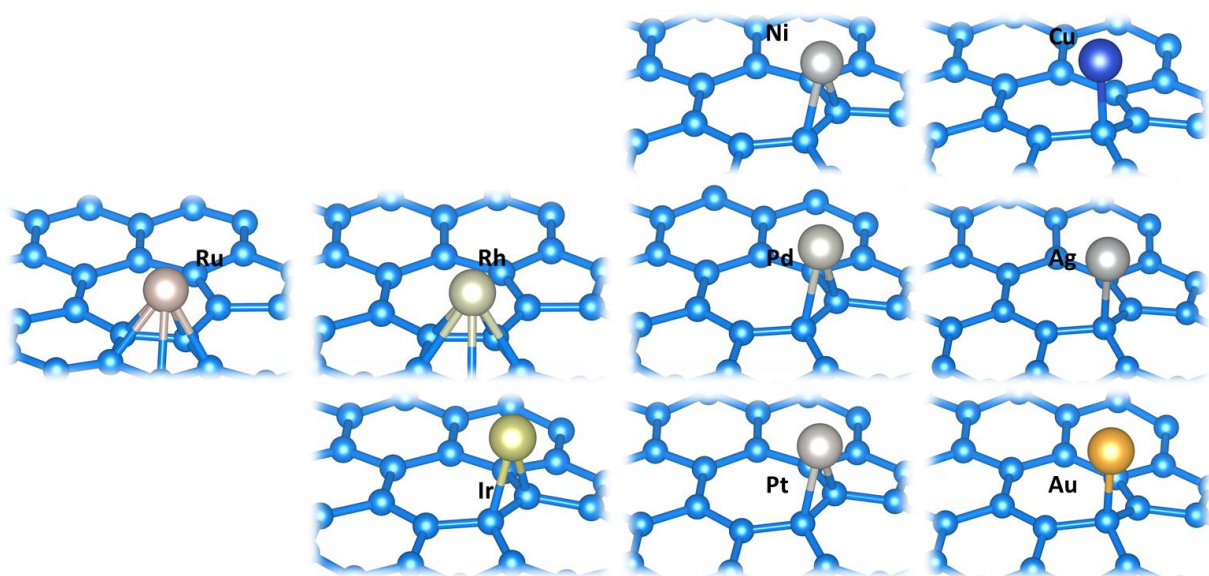

**Figure 6.** Optimized structures of selected d-elements and coinage metals adsorbed at the SW defect in the fully relaxed cell.

Obviously, the SW defect presents the center of altered reactivity of the graphene lattice, which is further affected by the mechanical deformation of the lattice. Thus, we have mapped the reactivity of the graphene lattice with the SW defect considering the fully relaxed cell (**Figure 7**) and the constrained cell scenario (**Figure S6**, Supplementary Information). The mapping was done considering the final position of the adsorbate atom upon the relaxation, as done in Ref. [49]. Briefly, the adsorption energy of the adatom at the site to which it relaxes was ascribed to the initial site in which it was placed. When all the sites in the cell were described with the corresponding binding energy of the relaxed structure, heatmaps were constructed by linear interpolation of these values. For both considered cases, constructed reactivity maps clearly show that the SW defect acts as an attractor for adsorbed metals. Also, the energy landscape of the SW-defective graphene lattice for the adsorption of studied metals is not exceptionally rough, and the energy difference was found to be the largest for Ru in the relaxed cell (0.95 eV). For the relaxed cell, reactivity maps are symmetric around the defect, while in the case of the constrained cell, the symmetry is still present, but it is lower compared to the fully relaxed cell. The obtained results suggest that coinage metals are likely highly mobile on the graphene surface with the SW defect and that their mobility can be triggered by applying mechanical deformation to the graphene sheet.

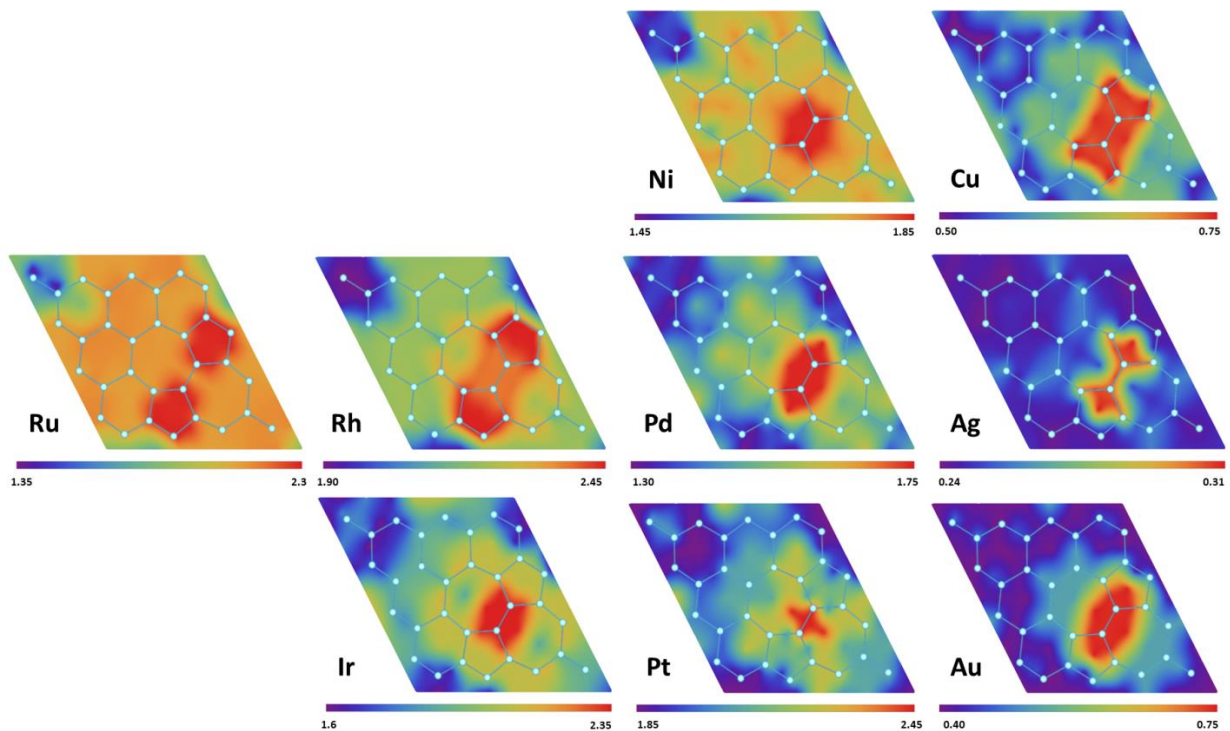

**Figure 7.** Reactivity maps of the SW defect in the fully relaxed cell towards selected d-elements and coinage metals. Color coding is provided below each map giving the value of the adsorbate binding energy in eV (blue – weaker bonding, red – stronger bonding).

Surprisingly, only a few metals exhibit non-zero magnetic moments ($M$) when adsorbed at the SW defect. Notably, Cu, Ag, and Ir exhibit non-zero magnetic moments in the relaxed cell configuration, while Cu, Ru, Rh, and Ir exhibit magnetization in the pristine cell calculations. In other cases, no magnetic moment is observed, while the d-states of atomic adsorbates are generally well localized in the energy window between −5 eV and the Fermi level (**Figure 8** and **Figure S7**, Supplementary Information). Subtle differences in the electronic structure can be observed when comparing the relaxed and constrained cell cases. Thus, we believe that the combination of single metal atom trapping at the SW sites, combined with the mechanical deformation of the graphene substrate, could be a plausible way to study the tuning of the reactivity of such formed metal single atoms (catalysts). This issue is, however, beyond the scope of the present work.

Ir was the only adsorbate that exhibited magnetization irrespective of the cell lattice (**Table 4**). **Figure 10** presents the orbital-resolved DOS and the spin density isosurface of Ir adatom on a fully relaxed graphene surface with the SW defect. Based on the spin density symmetry and the presented DOS, it can be concluded that the magnetism arises due to partially filled $d_{x2}$ orbitals, while hybridized $d_{xz}$ and $d_{z2}$ orbitals are mainly responsible for bonding with the graphene surface (bridge configuration).

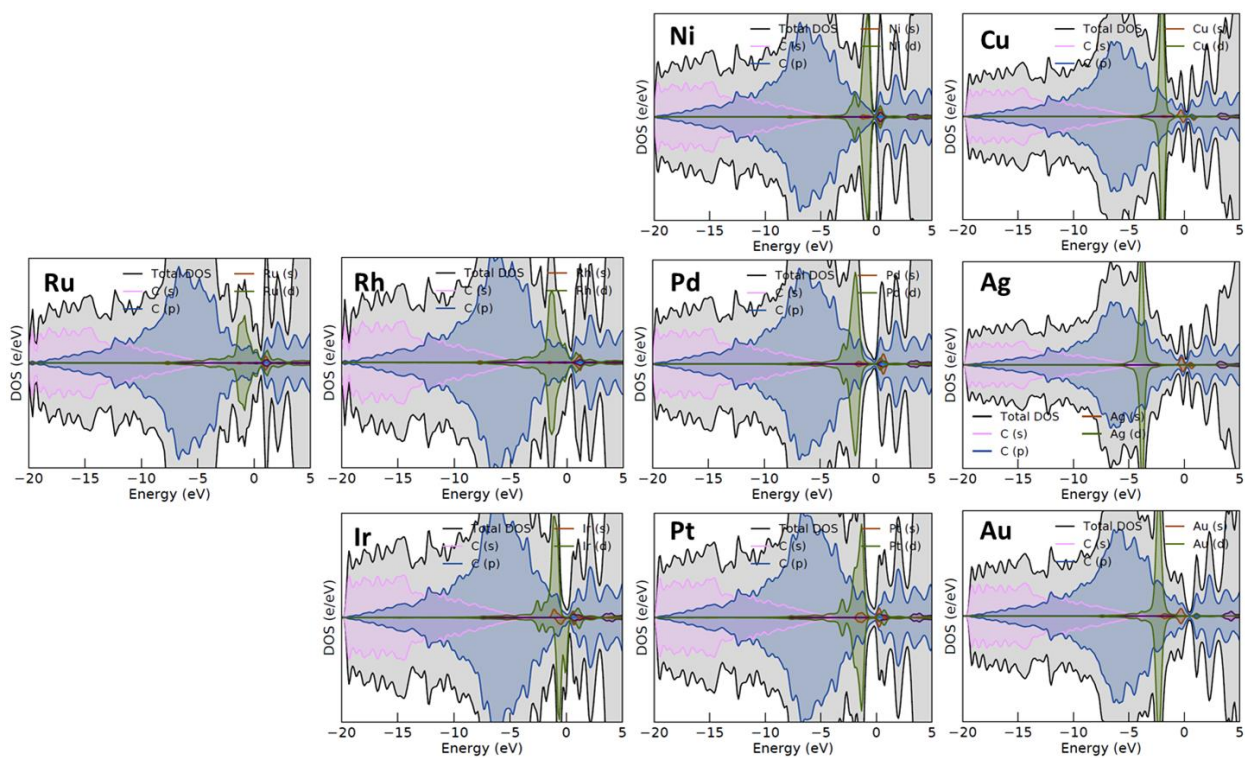

**Figure 8.** DOSes of selected d-elements and coinage metals adsorbed at the SW defect in the fully relaxed cell.

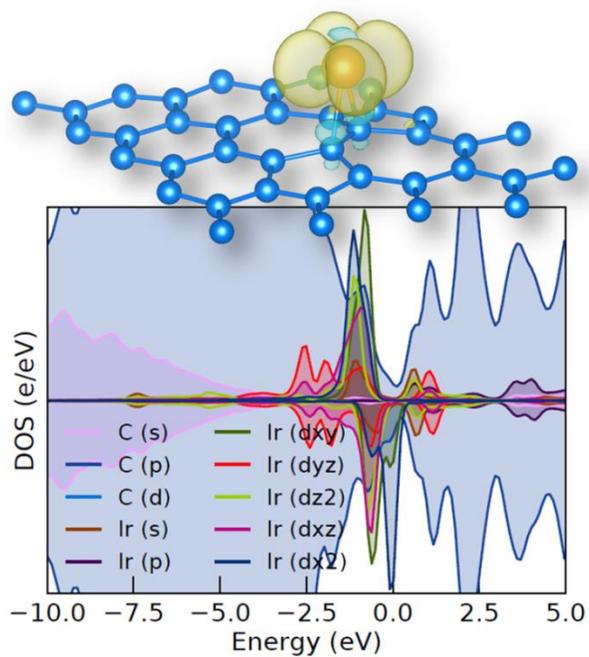

**Figure 10.** Spin density of Ir adatom at the SW defect (fully relaxed lattice, isosurface values $1.5×10^{-3}$ e Å$^{-3}$) and the corresponding lm-decomposed DOS.

## 3.6. SW-defected graphene vs. pristine graphene

When atomic adsorption on a SW-defective graphene surface (fully relaxed cell) is compared to the adsorption on the pristine graphene surface, it can be seen that in all the cases, the SW defect binds the studied atomic adsorbates stronger than the pristine graphene surface (**Figure 10**, a). Differences in the binding energies are up to ~1.4 eV, reaching the maximum for C and Si. As the SW defect was found to be an attractor for the studied adsorbates, it can be concluded that the introduction of the SW defect can be an effective way to fine-tune the reactivity of the graphene lattice. Even though there are no surface dangling bonds, the reactivity is improved compared to the pristine graphene surface. In contrast, the correlation between the binding energies on the defective and pristine surface is quite low (**Figure 10**, c). However, the binding on the SW defect is still much weaker compared to the single vacancy [23]. For example, none of the studied transition metal adsorbate binding energies were found to be above the cohesive energy of the corresponding bulk metal phase. Thus, the SW defect is not likely to be able to stabilize single atoms of these metals. However, these systems might have a high academic interest in studying fundamental aspects of graphene and graphene-supported single-atom catalysts' reactivity. Moreover, the mobility of these adatoms is generally high, although preferentially bonding is to the defect site.

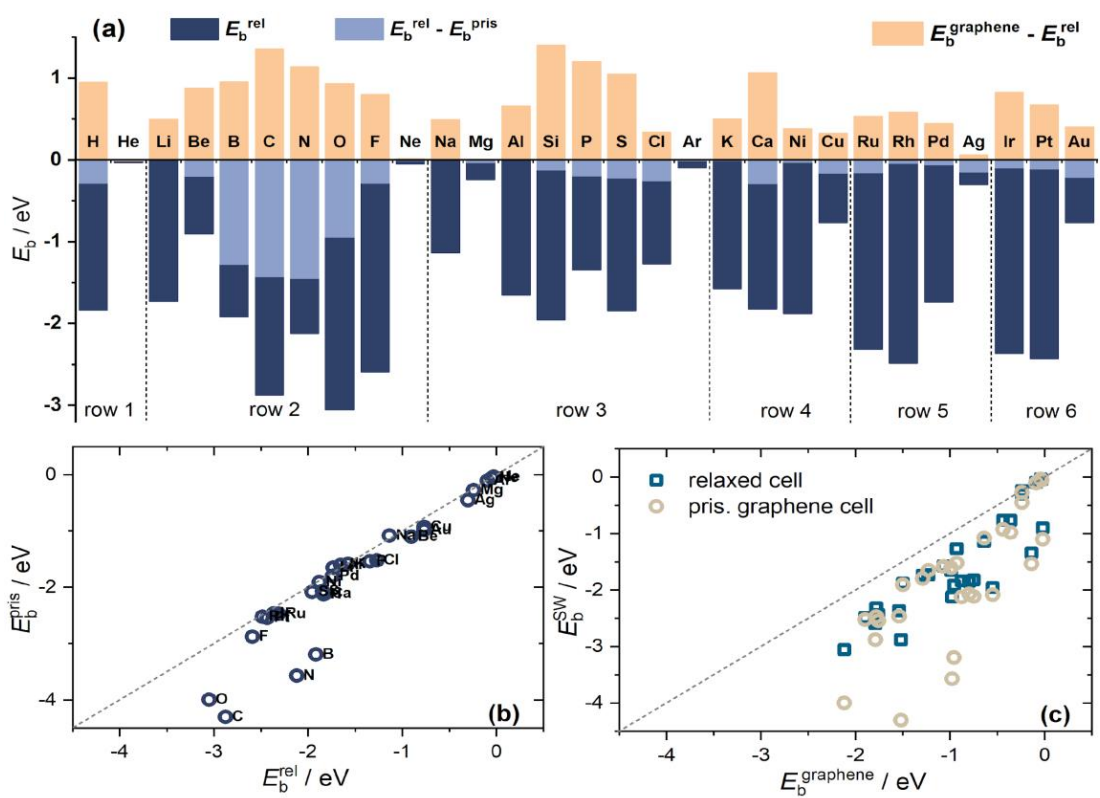

**Figure 10.** The overall adsorption trends: (a) Variation of calculated $E_b$ for selected elements in the fully relaxed cell along the rows of the PTE, compared to $E_b$ calculated for SW-graphene in the pristine cell and for the pristine graphene (data taken from Ref. [22] for PBE-D3), (a) correlation of $E_b$ calculated for adsorption on SW-graphene in fully relaxed cell and in pristine cell, (c) correlation of $E_b$ calculated for adsorption on SW-graphene in fully relaxed cell and on pristine graphene.

We also note that in the majority of cases, our calculations with the pristine graphene (constrained) cell led to more negative binding energies for the studied atomic adsorbates. A clear correlation exists between the binding energies calculated in the relaxed and constrained cell. B, C, N, and O are outliers with much more negative binding energy on the surface in the pristine graphene cell (**Figure 10**, b). Among these four elements, C, N, and O exhibit a particularly interesting behavior. Under mechanical deformation, they "destroy" the SW defect and cause the rearrangement of the carbon bond network and lead to the formation of the N- and O-doped defective graphene lattice.

## 4. Conclusions

Introducing the SW defect enhances the binding strength between atomic adsorbates and the graphene lattice. The SW defect acts as an attractor for studied atomic adsorbates. Despite the absence of surface dangling bonds, the reactivity of SW-defective graphene outperforms that of pristine graphene. However, the SW defect may not be suitable for stabilizing single atoms of certain transition metals. However, it holds potential for studying fundamental aspects of graphene-supported single-atom catalysts' reactivity. Thus, the defect engineering by introducing the SW defect, can be an effective strategy for fine-tuning the reactivity of graphene surface. Moreover, the presented results highlight the influence of lattice constraints on binding energies. Calculations with the pristine graphene (constrained) cell generally led to more negative binding energies for most atomic adsorbates compared to the fully relaxed cell. The behavior of elements B, C, N, and O deviates from this trend, exhibiting notably more negative binding energies in the pristine graphene cell. We find that under mechanical deformation, C, N, and O can "destroy" the SW defect, forming N- and O-doped defective graphene lattices. The presented work contributes to a deeper understanding of the reactivity of SW-defective graphene and its potential applications in catalysis and electrocatalysis. The stronger binding energies observed on the SW defect surface offer promise for designing novel graphene-based functional materials. However, further investigations are needed to explore the full potential of SW-defective graphene and its interactions with various atomic adsorbates.

## Acknowledgment

A.Z.J., A.S.D., and I.A.P. acknowledge the support provided by the Serbian Ministry of Science, Technological Development, and Innovation (451-03-47/2023-01/200146). The computations and data handling were enabled by resources provided by the National Academic Infrastructure for Supercomputing in Sweden (NAISS) at the NSC center of Linköping University, partially funded by the Swedish Research Council through grant agreement no. 2018-05973.

# Supplementary Information

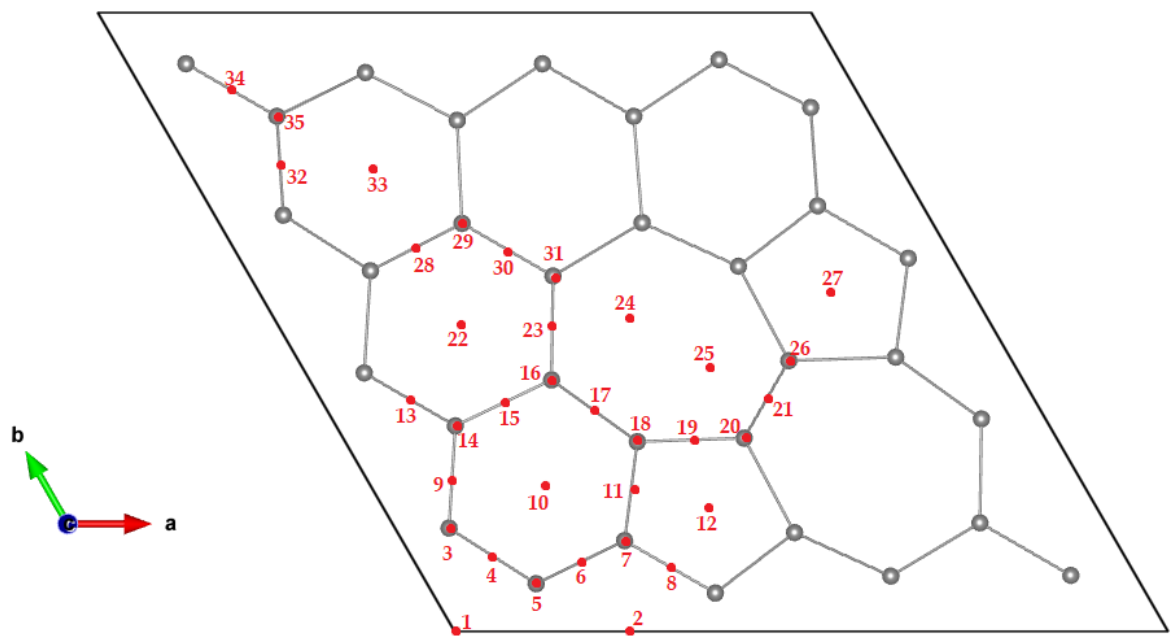

**Figure S1.** Selected adsorption site for atomic adsorption on the SW-graphene surface.

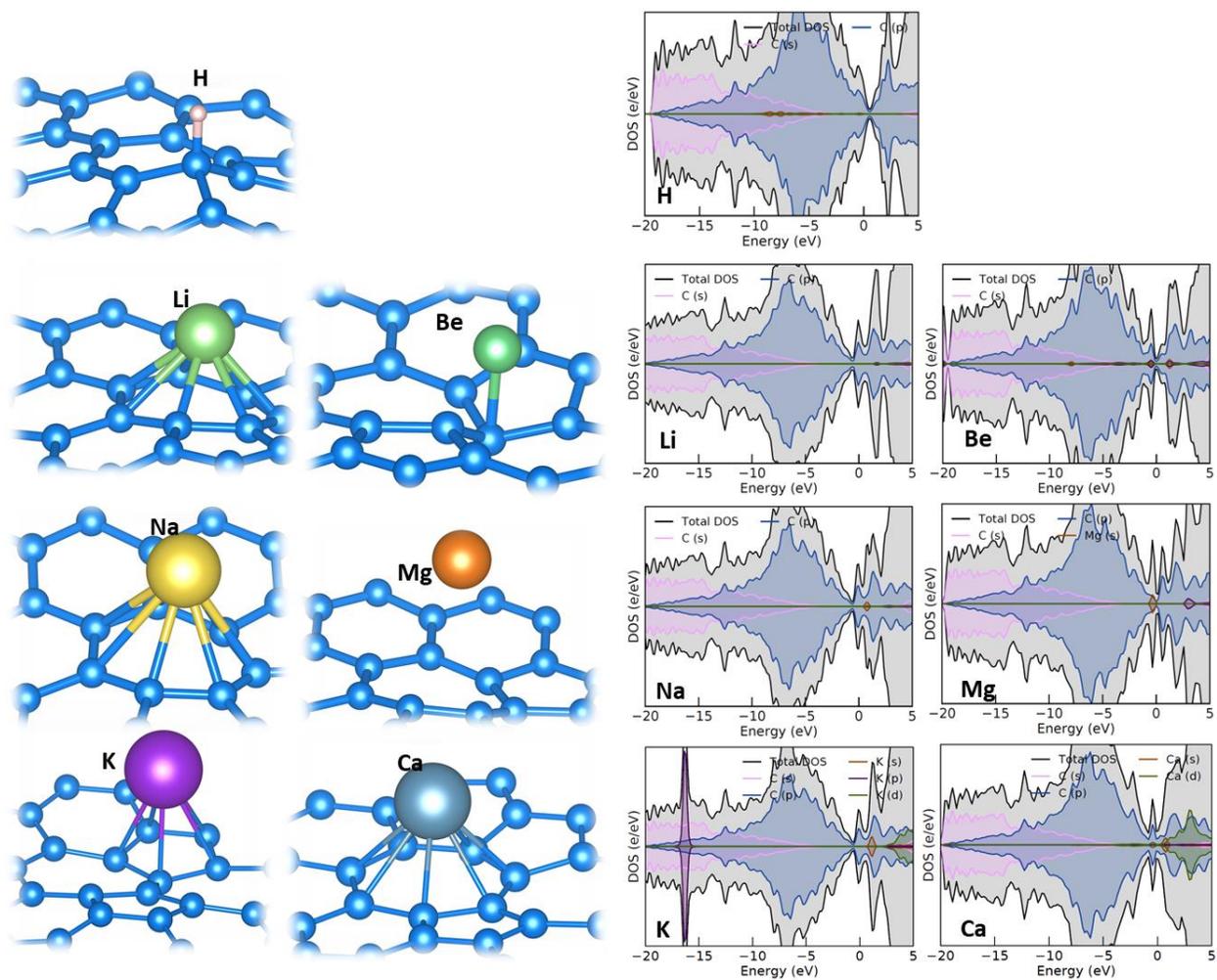

**Figure S2.** Optimized structures and DOSes for H, alkaline, and alkaline earth elements interacting with SW-graphene constrained to pristine graphene lattice.

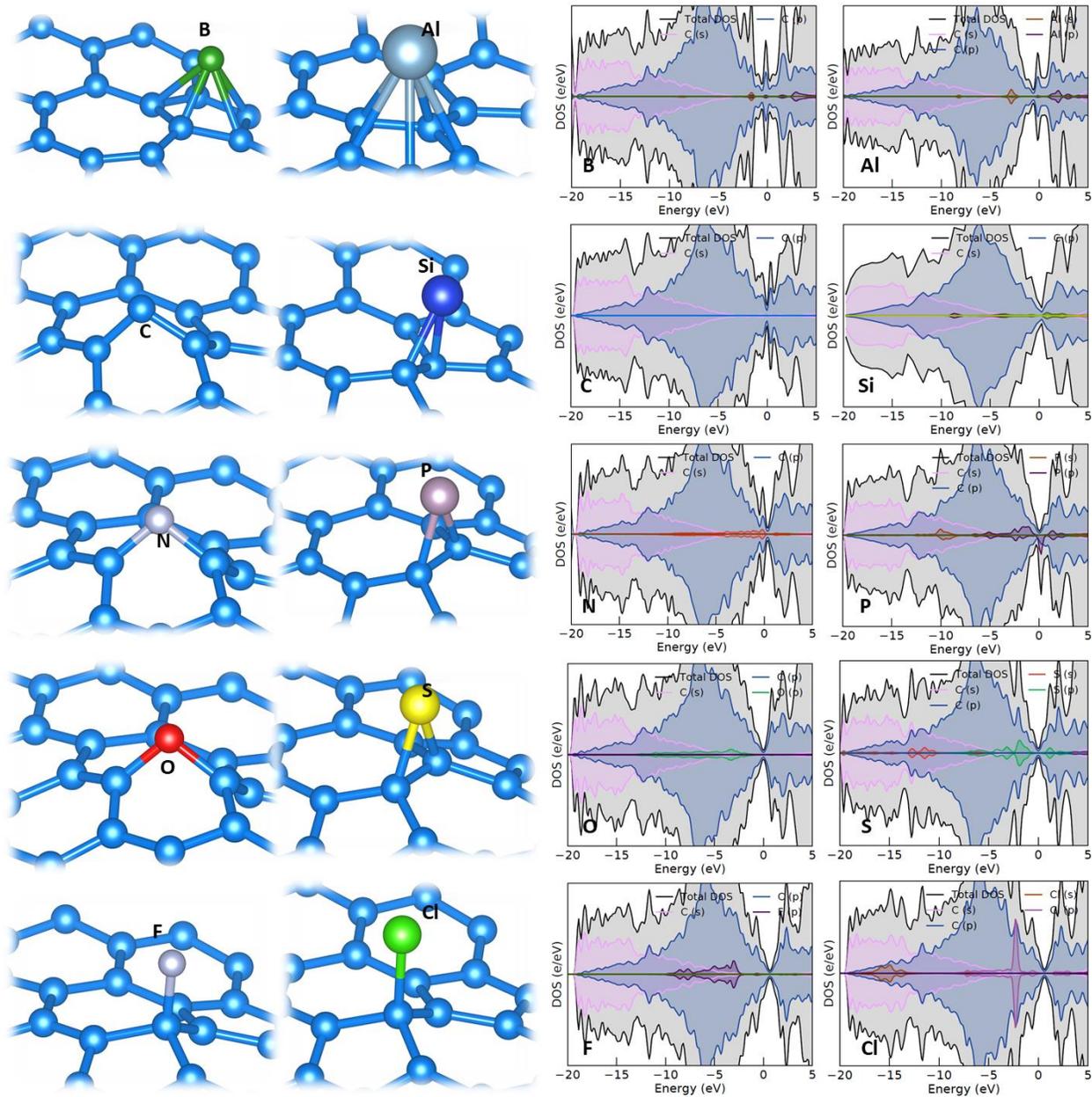

**Figure S3.** Optimized structures and DOSes for p-elements in rows 2 and 3 of the PTE at the SW defect in pristine graphene lattice.

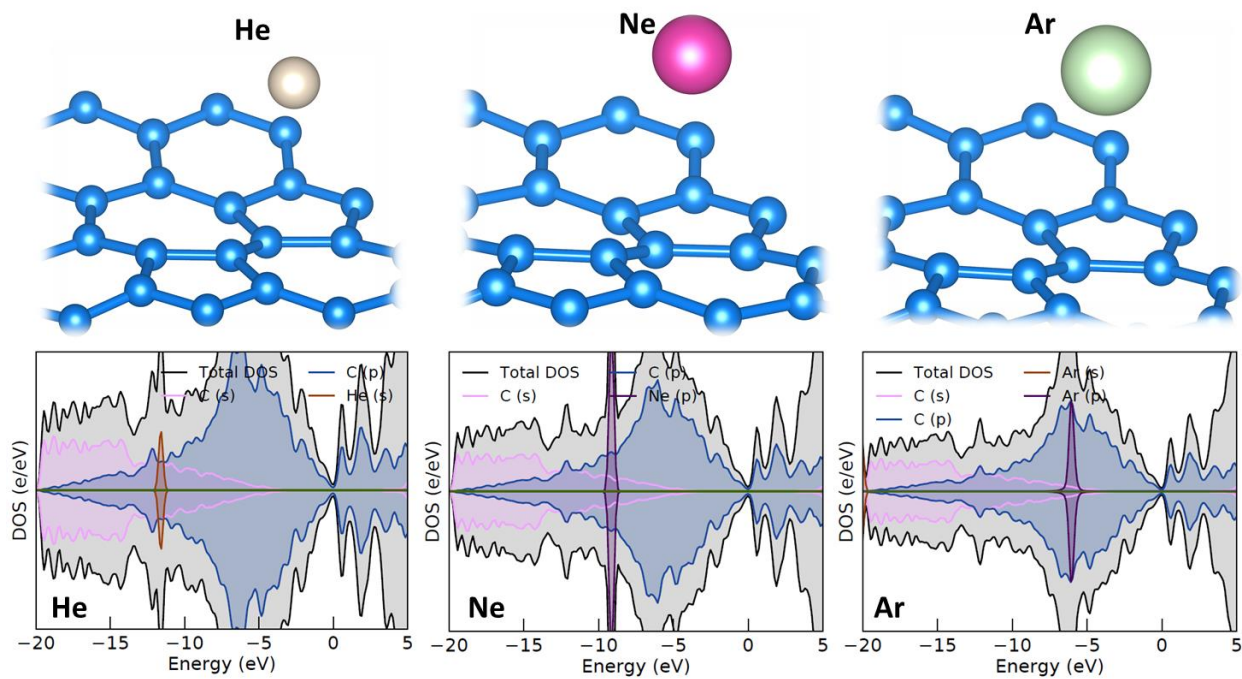

**Figure S4.** Optimized structures and DOSes of noble gases interacting with the SW defect in the pristine graphene lattice.

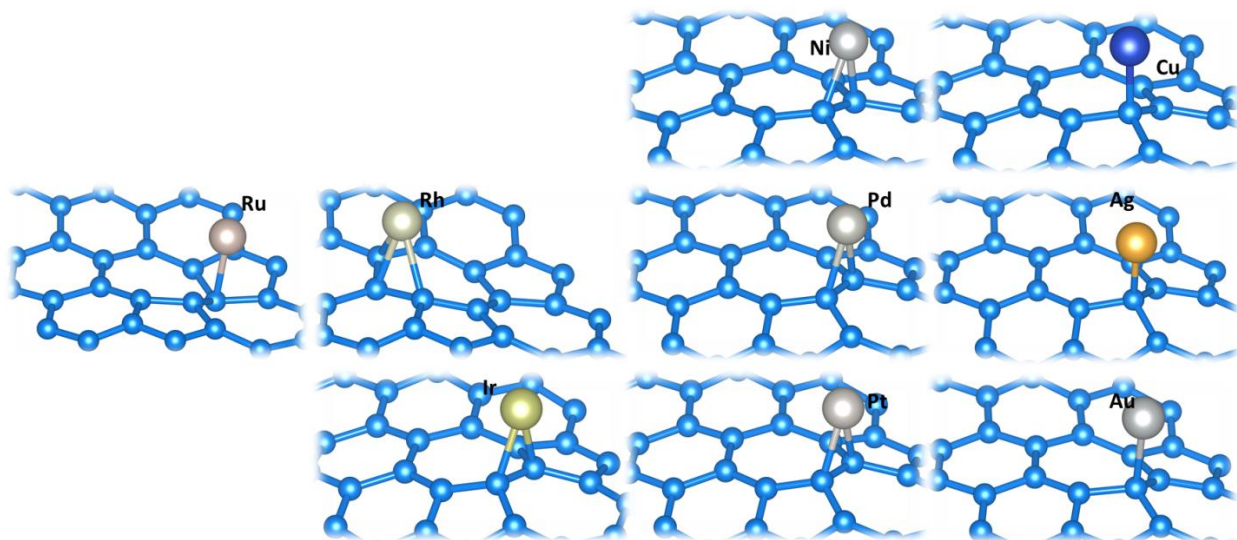

**Figure S5.** Optimized structures of selected d-elements and coinage metals adsorbed at the SW defect constrained to pristine graphene lattice.

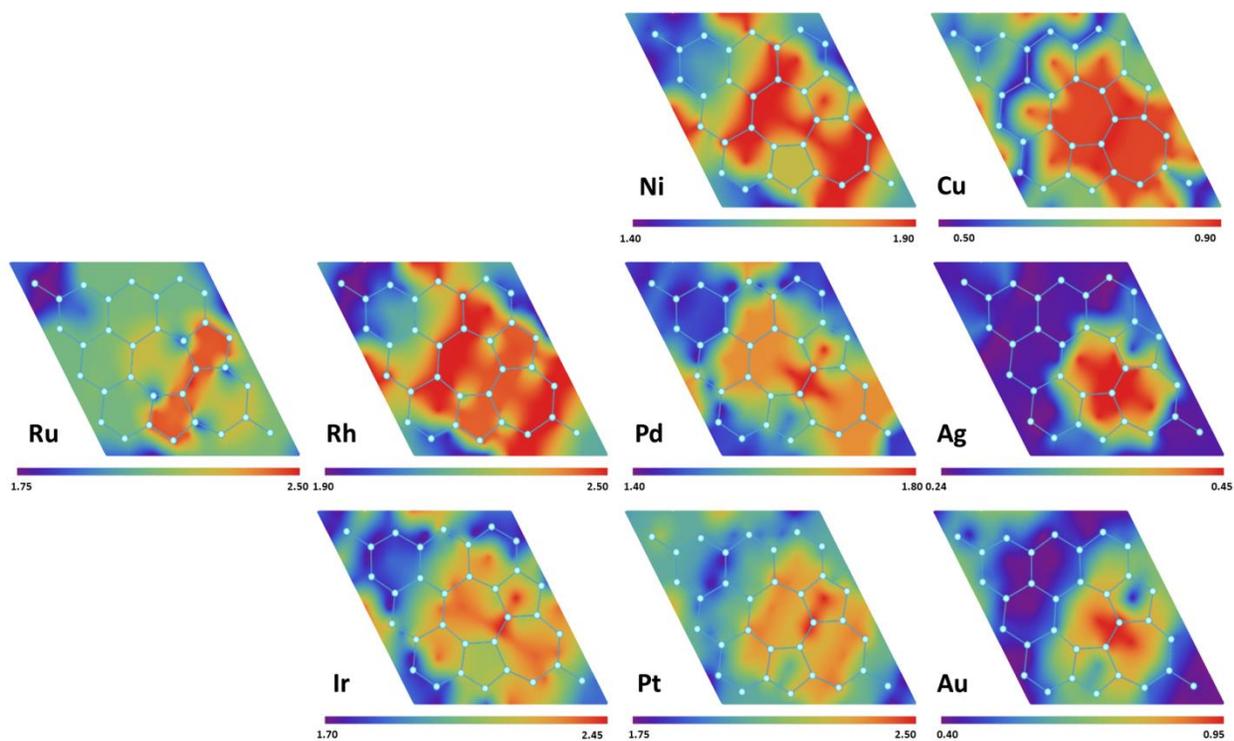

**Figure S6.** Reactivity maps of the SW defect in the pristine graphene lattice towards selected d-elements and coinage metals.

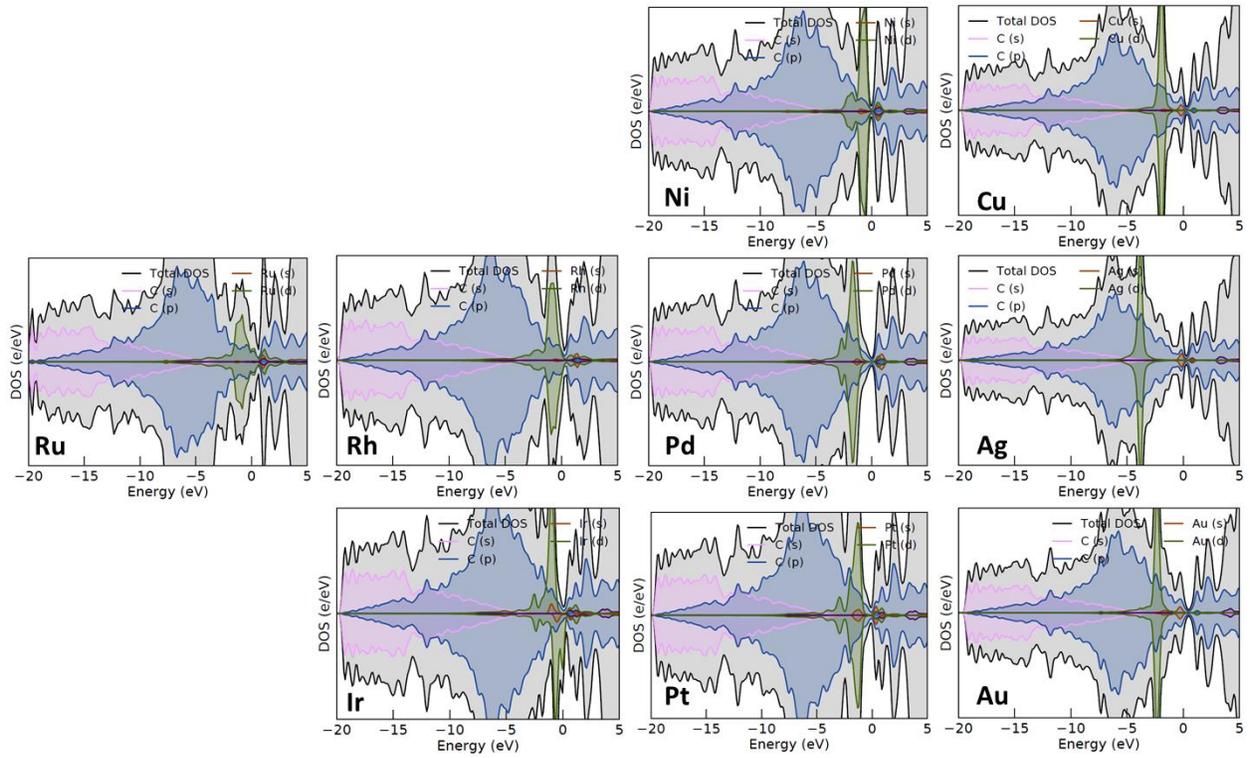

**Figure S7.** DOSes of selected d-elements and coinage metals adsorbed at the SW defect constrained to pristine graphene lattice.